\documentclass[preprint,12pt]{./elsarticle}

\usepackage{amssymb}

\usepackage{amsfonts,amsmath,amssymb,amsbsy,amsthm}

\usepackage{hyperref}
\usepackage{mathrsfs}
\usepackage{epstopdf}

\usepackage{graphicx}

\usepackage{subcaption}
\usepackage{multirow}

\usepackage{bbm} %indicator function

\begin{document}

\begin{frontmatter}

\title{A Critical Study of Cottenden \emph{et al.}'s \emph{An Analytical Model of the Motion
of a Conformable Sheet Over a General Convex Surface in the Presence of
Frictional Coupling}}

\author{Kavinda Jayawardana\corref{mycorrespondingauthor}}
\cortext[mycorrespondingauthor]{Corresponding author}
\ead{zcahe58@ucl.ac.uk}

\begin{abstract}
In our analysis, we show that what Cottenden \emph{et al.} \cite{cottenden2008development, cottenden2009analytical} and Cottenden \cite{cottenden2011multiscale} accomplish is the derivation of the ordinary capstan equation, and a solution to a dynamic membrane with both a zero-Poisson's ratio and a zero-mass density on a rigid right-circular cone. The authors states that the capstan equation holds true for an elastic obstacle, and thus, it can be used to calculate the coefficient of friction between human skin and fabrics. However, using data that we gathered from human trials, we show that this claim cannot be substantiated as it is unwise to use the capstan equation (i.e. belt-friction models in general) to calculate the friction between in-vivo skin and fabrics. This is due to the fact that such models assume a rigid foundation, while human soft-tissue is deformable, and thus, a portion of the applied force to the fabric is expended on deforming the soft-tissue, which in turn leads to the illusion of a higher coefficient of friction when using belt-friction models.
\end{abstract}


\end{frontmatter}

%% main text
\section{Introduction}
\label{S:1}

Consider a situation where two elastic bodies that are in contact with each other, where the contact area exhibits friction. A common area where modelling of such problems can be found in the field of tire manufacturing \cite{bowden1964friction, clark1981mechanics}. Now, consider the scenario where one of the elastic bodies is very thin and almost planar in a curvilinear sense, in comparison to the other body. Then the thin body can be approximated by a shell or a membrane, and such models can be used to model skin abrasion caused by fabrics as a result of friction. Need for valid modelling techniques are immensely important in fields such as field of incontinence associated dermatitis \cite{cottenden2009analytical}, sports related skin trauma \cite{ bergfeld1985trauma} and cosmetics \cite{asserin2000measurement}. It is documented that abrasion damage to human skin in cases such as the jogger's nipple \cite{levit1977jogger} and dermatitis from clothing and attire \cite{wilkinson1985dermatitis} are caused by repetitive movement of fabrics on skin, and in cases such as pressure ulcers \cite{maklebust2001pressure} and juvenile plantar dermatitis \cite{shrank1979aetiology}, friction may worsen the problem. In an attempt to  model such problem mathematically, Cottenden \emph{et al.} \cite{cottenden2009analytical} put forward a model in their publication \emph{An analytical model of the motion of a conformable sheet over a general convex surface in the presence of frictional coupling}. We show that, regardless of the authors' claims, what they derive is the ordinary capstan equation and a solution for a membrane with a zero-Poisson's ratio and a zero-mass density on a rigid right-circular cone. Cottenden \emph{et al.} \cite{cottenden2008development, cottenden2009analytical} and Cottenden \cite{cottenden2011multiscale} also claims that capstan equation can be used to calculate the friction between in-vivo skin and fabrics. With data gathered from human trials, we show that it is unwise to use \emph{belt-friction models} (e.g. capstan equation) to calculate the coefficient of friction between in-vivo skin and fabrics, as such models assume a rigid foundation, while human soft-tissue is elastic; thus, a portion of the applied force to the fabric is expended on deforming the soft-tissue, which in turn leads to the illusion of a higher coefficient of friction when using belt-friction models.

\section{Capstan Equation}

The capstan equation, otherwise known as Euler's equation of tension transmission, is the relationship governing the maximum applied-tension $T_\text{max}$ with respect to the minimum applied-tension $T_0$ of an elastic string wound around a rough cylinder. Thus, the governing equation can be express by the following equation,
\begin{align}
T_\text{max} = T_0 \exp(\mu_F\theta) ~, \label{CapstanEqn}
\end{align}
where $\theta$ is the contact angle and $\mu_F$ is the coefficient of friction. By \emph{string} we mean a one-dimensional elastic body, and \emph{rough} we mean that the contact area exhibits friction, where the coefficient of friction is the physical ratio of the magnitude of the shear force and the normal force between two contacting bodies. The capstan equation is the most perfect example of a \emph{belt-friction model}, which describes behaviour of a belt-like object moving over a rigid-obstacle subjected to friction \cite{rao2003engineering}. In engineering, the capstan equation describes a body under a load equilibrium involving friction between a rope and a wheel like circular object, and thus, it is widely used to analyse the tension transmission behaviour of cable-like bodies in contact with circular profiled surfaces, such as in rope rescue systems, marine cable applications, computer storage devices (electro-optical tracking systems), clutch or brake systems in vehicles, belt-pulley machine systems and fibre-reinforced composites \cite{jung2008capstan}.\\

For a rigorous study of the capstan equation (a membrane or a string over a rigid-obstacle) generalised to non-circular geometries, we refer the reader to chapter 2 of Jayawardana \cite{jayawardana2016mathematical}. There, as an example, the author present a solution to the capstan equation generalised for a rigid elliptical-prism, i.e. given a prism with a horizontal diameter of $2a$ and the vertical diameter of $2b$, parametrised by the map $\boldsymbol\sigma(x^1,\theta) = \boldsymbol(x^1, ~a\sin(\theta), ~b\cos (\theta) \boldsymbol)$, where $\theta$ is the  acute  angle that the vector $\boldsymbol (0, 0, 1\boldsymbol)_{\text{C}}$ makes with the vector $\boldsymbol (0, \varphi(\theta), 0\boldsymbol)$, and where $\varphi(\theta) = (b^2\sin^2(\theta)+a^2\cos^2(\theta))^\frac12$, the capstan equation takes the following form,
\begin{align}
T_{\text{elliptical}}(\theta) = T_0\exp\left(\mu_F \arctan\left(\frac{b}{a}\tan (\theta)\right)\right) \label{PrismEqn}~.
\end{align}
Note that equation (\ref{PrismEqn}) assumes that minimum applied-tension, $T_0$, is applied at $\theta=0$, and the vector brackets $\boldsymbol (\cdot \boldsymbol )$ implies that the vectors are in the Euclidean space and $\boldsymbol (\cdot \boldsymbol )_{\text{C}}$ implies that the vectors are in the curvilinear space.\\

\begin{figure}[!h]
\centering
\includegraphics[ width=0.75 \linewidth]{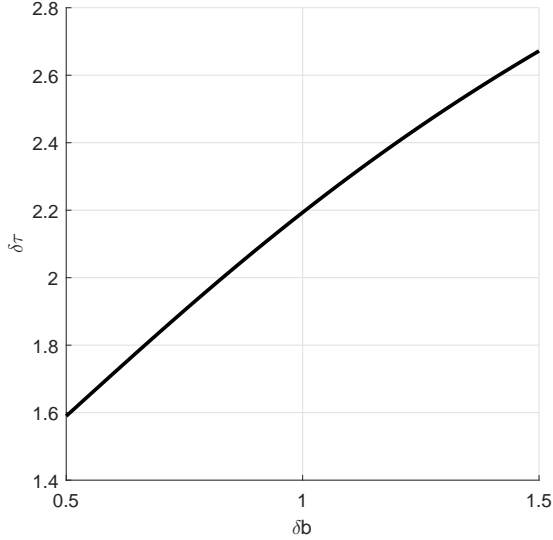}
\caption{Tension ratio against $\delta b$.\label{PrismEqnPlot2}}
\end{figure}

Equation (\ref{PrismEqn}) implies that the maximum applied-tension, $T_\text{max}$, is dependent on the mean curvature of the rigid prism. To investigate this matter further, assume that $T_0$ is applied at $\theta=-\frac14 \pi$ and $T_\text{max}$ applied at $\theta=\frac14 \pi$, and thus, equation (\ref{PrismEqn}) implies that
\begin{align}
\delta \tau = \exp\left(2\mu_F \arctan\left(\frac{b}{a}\right)\right) ~,\label{PrismEqnPlot}
\end{align}
where $\delta \tau = T_\text{max}/T_0$, which is defined as the \emph{tension ratio}. As the reader can see that for a fixed contact interval and a fixed coefficient friction, equation (\ref{PrismEqnPlot}) implies a non-constant tension ratio for varying $\delta b$, where $\delta b= b/a$. As the mean curvature of the prism is $ H(\theta) = \frac12 ab (\varphi(\theta))^{-3}$, one can see that the tension ratio is related to the mean curvature by the following relation,
\begin{align*}
\delta \tau = \exp\left(2\mu_F \arctan\left[\max_{\theta \in [-\frac14\pi, \frac14\pi]}(2a H(\theta),1)+\min_{\theta \in [-\frac14\pi, \frac14\pi]}(2a H(\theta),1)-1\right]\right) ~ .
\end{align*}

Figure \ref{PrismEqnPlot2} presents a visualisation of the tension ratio against $\delta b$ (which is analogous to the mean curvature), which is calculated with $\mu_F = \frac12$ and $\delta b \in [\frac12,  \frac32]$, and it  shows that, for a fixed contact interval, as $\delta b$ increases (i.e. as the mean curvature of the contact region increases), the tension ratio also increases. This is an intuitive result as the curvature of the contact region increases, the normal reaction force on the membrane (or the string) also increases, which in turn leads to a higher frictional force, and thus, a higher tension ratio. Now, this is a fascinating result as this effect cannot be observed with the ordinary capstan equation (\ref{CapstanEqn}).\\

For a rigorous study of belt friction models, we refer the reader to section 6.2 of Jayawardana \cite{jayawardana2016mathematical}, and for a rigorous study of the capstan equation generalised to elastic obstacles, we refer the reader to Konyukhov's \cite{ konyukhov2015contact}, and Konyukhov's and Izi's \cite {konyukhov2015introduction}.

\section{Cottenden \emph{et al.}'s Work}

Cottenden \emph{et al.} \cite{cottenden2009analytical} (the principal author is D. J. Cottenden) attempt to derive a mathematical model to analyse a frictionally coupled membrane (defined as a nonwoven sheet) on an elastic foundation (defined as a substrate) based on the research findings of D. J. Cottenden's PhD thesis \cite{cottenden2011multiscale}\footnote{\href{https://discovery.ucl.ac.uk/id/eprint/1301772/}{https://discovery.ucl.ac.uk/id/eprint/1301772/}}. It is assumed that friction (abrasion in the context of the publication) is the cause of some pressure ulcers in largely immobile patients, and abrasion due to friction contributes to the deterioration of skin health in incontinence pad users, especially in the presence of liquid. The current literature shows very little research in the area of frictional damage on skin due to fabrics, and thus, the authors' goal is to present a mathematical model to investigate this phenomenon in a purely geometrical setting. Thus, the authors propose a model for a general class of frictional interfaces, which includes those that obey Amontons' three laws of friction. \\

Cottenden \emph{et al.}'s  method  \cite{cottenden2009analytical} for calculating the kinetic frictional force induced on human skin due to nonwoven fabrics is as follows. The human body part in question is modelled as a homogeneous isotropic `convex surface [\emph{sic}]' \cite{cottenden2009analytical} (i.e. the substrate) and the nonwoven fabric is modelled as an isotropic membrane (i.e. the nonwoven sheet). The goal is to find the stresses acting on the nonwoven sheet, including determining the friction acting on the substrate. The contact region between the fabric and the skin is defined as `An \emph{Instantaneous Isotropic Interface}, [which] is an interface composed of a pair of surfaces which have no intrinsically preferred directions and no directional memory effects, so that the frictional force acts in the opposite direction to the \emph{current} relative velocity vector ... or to the sum of \emph{current} applied forces acting to initiate motion ...' (see section 2.2 of Cottenden \emph{et al.} \cite{cottenden2009analytical}): this simply implies that the contact region is isotropic and nothing more. Also, consider the contact body in question: it is modelled as a surface, i.e. a two-dimensional manifold. However, in reality, it must be modelled  as a three-dimensional object as a two-dimensional object cannot describe the elastic properties of a fully three-dimensional object such as a human body part, as a two-dimensional surface has measure zero in three-dimensional space (for an introduction on measure theory, please consult chapter 1 of Kolmogorov \emph{et al.} \cite{kolmogorov1961measure}); Unless suitable assumptions are made as one finds in shell theory (for a comprehensive mathematical study of the shell theory, please consult sections B of Ciarlet \cite{Ciarlet}); however, this is not what the authors are considering. Now, consider the authors' statement regarding the modelling assumptions carefully, particularly the term `convex surface'. The authors definition of convexity is $\boldsymbol \eta\cdot\boldsymbol\nabla_{\!\text{E}} \hat{\textbf{N}} \cdot \boldsymbol \eta \geq 0$ (see section 3.1 of Cottenden \emph{et al.} \cite{cottenden2009analytical}), where $\hat{\textbf{N}}$ and $\boldsymbol \eta$ are unit normal and tangential vectors surface respectively. However, the authors' definition is erroneous. Convexity has a very precise mathematical definition, i.e. we say the functional $f:X \to \mathbb{R}$ is convex, if $f(t x + (1-t)y)\leq t f(x) + (1-t)f(y)$, $\forall~t\in [0,1]$ and $\forall~x,y \in X$ (for more on the definition of convexity, please see chapter 1 of Badiale and Serra \cite{badiale2010semilinear}). Also, the very idea of a convex surface is nonsensical as definition of convexity is only applicable to functionals. A simple example of a convex functional is $\exp(\cdot):\mathbb R \to \mathbb R_{>0}$. One is left to assume that what the authors mean by convexity is surfaces (i.e. manifolds) with a positive mean-curvature. For more on elementary differential geometry, please consult do Carmo \cite{do2009differential} or Lipschutz \cite{lipschutz1969schaum}.\\

In their analysis, the authors defines a membrane with the following property, `always drapes, following the substrate surface without tearing or puckering' (see section 2.1 of \cite{ cottenden2009analytical}). The authors' definition is erroneous, as one cannot guarantee that the given property will hold for a flat-membrane (in a Euclidean sense) over an arbitrary curved surfaces. To illustrate the flaw, consider a flat elastic-membrane (i.e. a film) over a rigid sphere. The only way one can keep the membrane perfectly in contact with the sphere in a two-dimensional region with nonzero measure is by deforming the membrane by applying appropriate boundary stresses and or external loadings. Otherwise, the membrane only makes contact with the sphere at a single point or a line. Also, the authors do not specify whether the membrane is elastic or not. One is left to assume that the membrane is elastic as the proposed frame work does not acknowledge plastic deformations. Note that the authors never referred to their nonwoven sheet as a membrane, but a membrane (or a film) is the closest mathematical definition for modelling such objects. Please consult  chapters 4 and 5 of Ciarlet \cite{Ciarlet} or chapter 7 of Libai and Simmonds \cite{libai2005nonlinear} for a comprehensive analysis on the theory of membranes.\\

To find the stresses acting on the membrane, consider Cauchy's momentum equation in the Euclidean space, which the authors define as follows,
\begin{align}
\boldsymbol \nabla_{\!\text{E}} \cdot \textbf T + \textbf f = \rho \ddot{\boldsymbol \chi}~, \label{cauchy10}
\end{align}
where $\textbf T$ is Cauchy's stress tensor, $\textbf f = \textbf f (\textbf T,\boldsymbol \nabla_{\!\text{E}}\textbf T)$ is the force density field and $\rho$ is the material mass density of the membrane, $\boldsymbol \nabla_{\!\text{E}}$ is the Euclidean differential operator and $\boldsymbol \chi$ is given as a `time-dependent deformation function mapping the positions of points in their undeformed \emph{reference} configuration to their deformed positions and the superposed double dot denotes a double material description time derivative'  (see section 2.1 of Cottenden \emph{et al.} \cite{cottenden2009analytical}). It is unclear what $\boldsymbol \chi$ represent from the authors' definition, whether it is the displacement field of the membrane or some time-dependent mapping from one manifold to another. If the latter is true, then equation (\ref{cauchy10}) has a very different meaning, i.e. it means that the space is dependent of time, and such problems are encountered in the field of cosmology. If the reader consults section 5.4 of Cottenden's thesis \cite{cottenden2011multiscale}, then it becomes evident that $\boldsymbol \chi$ is, indeed, a time dependent map. However, if one consults Cottenden \emph{et al.} \cite{cottenden2008development, cottenden2009analytical} and Cottenden \cite{cottenden2011multiscale}, then one concludes that the authors do not put forward the framework to handle the 3+1 decomposition in general relativity, with any mathematical rigour. If the reader is interested in the 3+1 formalism in general relativity, then  please consult the publications \cite{griffiths2009exact, stephani2003exact, wald2010general} or Dr J. A. V. Kroon (QMUL) on his LTCC lecture notes on \emph{Problems of General Relativity}\footnote{ \href{http://www.maths.qmul.ac.uk/~jav/LTCCmaterial/LTCCNotes.pdf}{ http://www.maths.qmul.ac.uk/$\sim$jav/LTCC.htm}}, where the reader can find an extraordinary solution for two merging black-holes (i.e. the Brill-Lindquist solution).\\

Assuming that the foundation is static and rigid, and the mass density of the membrane is negligible, i.e. $\rho \approx 0$, the authors state that Cauchy's momentum equation (\ref{cauchy10}) can be expressed as
\begin{align}
\textbf P_{s}\cdot (\boldsymbol \nabla_{\!\text{E}} \cdot \textbf T) + \textbf P_{s} \cdot \textbf f & = \boldsymbol 0~,\label{dave1}\\
- (\boldsymbol \nabla_{\!\text{E}}\hat{\textbf N}) : \textbf T + \hat{\textbf N} \cdot \textbf f & = 0~, \label{dave2}
\end{align}
where $\textbf P_{s}$ projection matrix to the substrate (the explicit form is not defined by the authors), $\hat{\textbf N}$ is the unit normal to the surface, and $\cdot$ and $:$ are defined as a contraction and a double contraction in the Euclidean space respectively. Although it is not explicitly defined, one must assume that the authors use the fact that membranes cannot support any normal stresses, i.e. $\hat{\textbf N} \cdot \textbf T =\boldsymbol 0$, to obtain equation (\ref{dave2}). The authors give equations (\ref{dave1}) and (\ref{dave2}) as the state of the `general case' of the problem. However, their assertion cannot hold as the system is underdetermined. Consider the vector $\textbf f$, which consists of three unknowns. Also, consider the tensor $\textbf T,$ which is a symmetric tensor with six unknowns. Thus, using the condition $\hat{\textbf N} \cdot \textbf T = \boldsymbol 0$ the number of unknowns can be reduced by three: leaving six remaining unknowns. Now, direct one's attention to equations (\ref{dave1}) and (\ref{dave2}), which provide three additional equations. Thus, one comes to the conclusion that one has an underdetermined system, with three equations and six unknowns. Furthermore, there is no description of any boundary conditions for the `general case', which are essential in obtaining a unique solution.\\

The derivation of Cottenden \emph{et al.}'s  governing equations \cite{cottenden2011multiscale} can be found on section 2.2 to 2.4 of the publication. Upon examination, the reader may find that the methods put forward by the authors' are inconsistent with mathematical elasticity, contact mechanics and differential geometry. Thus, we refer reader to Kikuchi and Oden \cite{Kikuchi} to see how to model friction with mathematical competence and to show how incredibly difficult modelling such problems are. We further refer the reader to Ciarlet \cite{Ciarlet} to see how to model mathematical elasticity in a differential geometry setting with mathematical rigour.\\

To find explicit solutions, the authors direct their attention to only `\textbf{surfa-ces that are isomorphic to the plane; that is, those which have the same first fundamental form as the plane; the identity matrix in the case of plane Cartesian coordinates.} [\emph{sic}]' \cite{cottenden2009analytical}. Found in section 4.1 of Cottenden \emph{et al.} \cite{cottenden2009analytical}, this is the basis for their entire publication (also the basis for Cottenden's \cite{cottenden2011multiscale} thesis). However, the authors' statement is nonsensical. An \emph{isomorphism} (preserves form) is at least a homomorphism, i.e. there exists at least a continuous bijective mapping, whose inverse is also continuous, between the two manifolds in question \cite{do2009differential, lipschutz1969schaum}. Thus, a surface that is isomorphic to the plane simply implies that there exists a continuous bijective map, with a continuous inverse, between the surface in question and the Euclidean plane, and it does not automatically guarantee that the surface have the same metric as the Euclidean plane under the given map. The latter part of the authors' statement is clearly describing surfaces that are \emph{isometric} (preserves distance) to the Euclidean plane, i.e. surfaces with a zero-Gaussian curvature. However, the statement is still erroneous as being a surface that is isometric to the Euclidean plane does not guarantee that the surface have the same metric as the Euclidean plane. Being isometric to the Euclidean plane simply implies that, if $f:U\subset \mathbb{R}^2 \to W \subset \textbf{E}^3$ is a 2D manifold that is isometric to the Euclidean plane, then there exists a map $g :V\subset \mathbb{R}^2 \to U\subset \mathbb{R}^2$ such that the first fundamental form of the isometry $f\circ g :V\subset \mathbb{R}^2 \to W \subset \textbf{E}^3$ is the $2 \times 2$ identity matrix \cite{do2009differential, lipschutz1969schaum}. One is left to assume that the surfaces that are in question by the authors belongs a subgroup of surfaces with zero-Gaussian curvature that has the same metric as the Euclidean plane with respect to the authors' coordinate system, i.e. cylinders and right-circular cones: as one later see that these are the only possible manifolds that can generate any valid solutions. Note that Cottenden \cite{cottenden2011multiscale} accredits Pressley \cite{pressley2010elementary} for his differential geometry results. However, Pressley's publication \cite{pressley2010elementary} is a widely accepted and verified mathematical publication in the field of differential geometry, and it  does not contain such provably false statements as given by Cottenden \cite{cottenden2011multiscale}.\\

Now, consider the equation
\begin{align}
\textbf P_{s} \cdot \textbf f + \mu_d(\hat{\textbf N} \cdot \textbf f)\dot{\boldsymbol \chi} = 0~, \label{dave3}
\end{align}
which the authors define as Amontons' law friction, where $\mu_d$ is the coefficient of dynamic friction and $\dot{\boldsymbol \chi}$  is the relative velocity vector between the membrane and the foundation. The inclusion of the two equations implied by condition (\ref{dave3}) still does not guarantee that the system is fully determined, as the system requires one more equation to be fully determined.\\

Now, assume that one is dealing with a rectangular membrane whose orthogonal axis defined by the coordinates $\boldsymbol(x, y\boldsymbol)$, where $y$ defines the longer dimension, that is placed over a surface defined by the regular map $\boldsymbol \sigma$. Also, assume that Poisson's ratio of the membrane is zero to prevent any lateral contractions due to positive tensile strain. To reduce the complexity, the authors modify the problem by letting $\dot{\boldsymbol \chi}$ be parallel to $\boldsymbol \sigma_{\!,y}$. Also, by applying a boundary stress of $T_0$ at some point $\phi_1$ whilst applying a even greater stress at $\phi_2$ so that $T_{yy} (y)$ is an increasing function in $y$, where $\phi_1$ and $\phi_2$ are angles of contact such that $\phi_1<\phi_2$. Due to zero-Poisson's ratio and the boundary conditions, one finds $T_{xx} = 0$, $T_{xy} =0$, where $T_{ij}$ are stress tensor components. Thus, the governing equations finally reduce to a fully determined system, i.e. is only now the number of unknowns equals to the number of governing equations. Therefore, one must understand that having zero-Gaussian curvature and zero-Poisson's ratio is a necessity for this model, and it is not some useful tool for deriving explicit equations as stated by the authors. Upon integrating equation (\ref{dave3}), under the specified boundary conditions, one finds solutions of the following form (see equation 4.4 of Cottenden \emph{et al.} \cite{cottenden2009analytical}),
\begin{align}
T_{yy}(y) = T_0 \exp\left( \mu_d \int^y | C_{yy} (\eta)| ~d \eta \right)~, \label{deadgive}
\end{align}
where $C_{\alpha\beta}= F_{I \alpha \gamma }^{-1} F_{II \gamma \delta }  F_{I \beta  \delta }^{-1}$ is defined as the curvature tensor, ${\boldsymbol F}_{\!I}$ is the first fundamental form tensor and $F_{II yy}$ is the only nonzero component of the second fundamental form tensor. However, equation (\ref{deadgive}) is erroneous. This is because, whatever is inside the $\exp(\cdot)$ term must be non-dimensional, but this is not the case with equation (\ref{deadgive}). To illustrate this flaw, let $L$ be an intrinsic Euclidean length scale and $\ell$ be an intrinsic length scale of the curvilinear coordinate $y$. Now, with the definition of $C_{yy}$ (see equation 3.3 of Cottenden \emph{et al.} \cite{cottenden2009analytical}), one finds that the length scale inside the $\exp(\cdot)$ term in equation (\ref{deadgive}) to be $ {(\ell/L)}^3$. Given that $y=\theta$ (i.e. the contact angle, which is dimensionless), one finds that the length scale inside the $\exp(\cdot)$ term to be $ L^{-3}$, which is not mathematically viable. This flaw of the authors' work is a result of not correctly following tensor contraction rules, and not discerning between covariant and contravariant tensors (see sections 3.2 and 4.1 of Cottenden \emph{et al.} \cite{cottenden2009analytical}). For elementary tensor calculus, please consult Kay \cite{kay1988schaum}.\\

If one assumes a 2D manifold that is \underline{\emph{\textbf{isometric}}} to the Euclidean plane, has a positive mean-curvature (with respect to the unit-outward normal) and whose first fundamental form tensor is diagonal (after a change of coordinates or otherwise), and should one follow the derivation with mathematical rigour, then one finds a solution of the following form,
\begin{align}
T_{y}^y(y) = T_0 \exp\left( -\mu_d \int^y \sqrt{F_{I yy } (\eta)} F_{IIy}^{~~y} (\eta)  ~d \eta \right)~, \label{deadgive2}
\end{align}
where a rigorous derivation can be fond in chapter 2 of Jayawardana \cite{jayawardana2016mathematical}. As the reader can see that unlike equation (\ref{deadgive}), no dimension violations can be possible with equation (\ref{deadgive2}).\\

To find the explicit solution for the general \emph{prism} case, the authors present the following map (see section 4.2 of Cottenden \emph{et al.} \cite{cottenden2009analytical})
\begin{align}
\boldsymbol\sigma(x,y) &= \boldsymbol(R(\phi)\cos(\phi), ~R(\phi)\sin(\phi), ~x\cos(\zeta) + y\sin(\zeta) \boldsymbol)~, \label{dave6}
\end{align}
where
\begin{align*}
d\phi  &= \frac{\cos(\zeta) dy - \sin(\zeta) dx}{\sqrt{R(\phi)^2 + (R^\prime(\phi))^2}} ~. 
\end{align*}
From the authors'  definition, $\zeta$ appears to be the acute angle that the vector $\hat{\boldsymbol \sigma}_{\!,y}$ makes with the vector $\hat {\boldsymbol \sigma}_{\!,\phi}$, and $R$ and $\phi$ appear to be the radius of curvature and the angle of the centre of rotation respectively. One can clearly see that map (\ref{dave6}) is only valid for cylinders (i.e. prisms with a constant radius) as it must have the same metric as the Euclidean plane. To be precise,
\begin{align*}
\boldsymbol F_{I}=
&\begin{pmatrix}
{\boldsymbol \sigma}_{\!,x}\cdot {\boldsymbol \sigma}_{\!,x} & {\boldsymbol \sigma}_{\!,x}\cdot {\boldsymbol \sigma}_{\!,y}\\
{\boldsymbol \sigma}_{\!,x}\cdot {\boldsymbol \sigma}_{\!,y} & {\boldsymbol \sigma}_{\!,y}\cdot {\boldsymbol \sigma}_{\!,y}
\end{pmatrix}\\
= &[R(\phi)^2 + (R^\prime(\phi))^2] \begin{pmatrix}
(\phi_{,x})^2 + \cos^2(\zeta)  & \phi_{,x}\phi_{,y} + \sin(\zeta)\cos(\zeta)\\
\phi_{,x}\phi_{,y} + \sin(\zeta)\cos(\zeta) & (\phi_{,y})^2 + \sin^2(\zeta)
\end{pmatrix}\\
= &\begin{pmatrix}
1 &0\\
0 & 1
\end{pmatrix} ~ ,
\end{align*}
which, in turn, implies the following,
\begin{align*}
\phi_{,x} &= - \frac{\sin(\zeta)}{\sqrt{R(\phi)^2 + (R^\prime(\phi))^2}}~,\\
\phi_{,y} &= \frac{\cos(\zeta)}{\sqrt{R(\phi)^2 + (R^\prime(\phi))^2}}~,
\end{align*}
and thus, the following,
\begin{align}
(\phi_{,x})^2  + (\phi_{,y})^2  &= \frac{1}{R(\phi)^2 + (R^\prime(\phi))^2}~. \label{mapcyl}
\end{align}
As $\phi$ is a real function (i.e. not complex) and noting that equation (\ref{mapcyl}) must hold for all $x$ and $y$, one finds that $R^\prime = 0$, and thus, equation (\ref{dave6}) reduces to the following,
\begin{align*}
\boldsymbol\sigma(x,y) &= \boldsymbol(c\cos(\phi), ~c\sin(\phi), ~x\cos(\zeta) + y\sin(\zeta) \boldsymbol)~,\\
\phi(x,y)  &= \frac{1}{c}[\cos(\zeta) y - \sin(\zeta) x]~,
\end{align*}
which represents a parametrisation of a cylinder, where $c$ is a positive constant (and $R=c$). Now, given that a solution exists in the interval $[\phi_1,\phi_2]$, the authors state that the solution can be expressed as follows (see equation 4.7 of Cottenden \emph{et al.} \cite{cottenden2009analytical}),
\begin{align}
T_{yy}(\phi_2) = T_0\exp\left(\mu_d \cos(\zeta)\left [\phi - \arctan\left(\frac{R(\phi)_{,\phi}}{R(\phi)}\right)\right]\big|^{\phi_2}_{\phi_1}\right)~. \label{dave4}
\end{align}

\begin{figure}[!h]
\centering
\includegraphics[trim = 2cm 0cm 1cm 0cm, clip = true, width=1\linewidth]{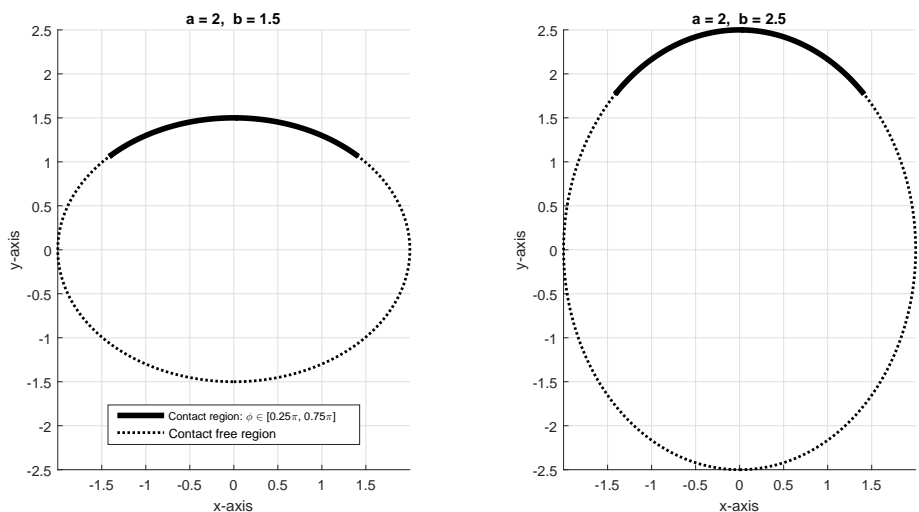}
\caption{Cross sections of elliptical-prisms (and elliptical-cones for the $z=1$ case). \label{treck1} }
\end{figure}

Regardless of the authors' claims, solution (\ref{dave4}) is only valid for cylinders, i.e. the capstan equation (\ref{CapstanEqn}). To see why equations (\ref{dave6}) and (\ref{dave4}) are not valid for general prisms, one only needs to consider an example with noncircular cross section. If the reader wishes to, then consider a positively-oriented elliptical-prism (for the $\zeta= 0$ case) that is defined by the map 
\begin{align}
\boldsymbol\sigma(\phi,z) = \boldsymbol(a\cos(\phi), ~b\sin(\phi), ~z\boldsymbol)~, \label{ellipX}
\end{align}
where $z \in \mathbb R$ and $a,b>0$, and let $\phi \in [\frac14\pi,\frac34\pi]$ be the contact interval (see figure \ref{treck1}, and see equation (\ref{PrismEqn}) for the capstan equation for an elliptical-prism). Now, the reader can  see that both map (\ref{dave6}) and solution (\ref{dave4}) are no longer valid for the elliptical-prism mapping (\ref{ellipX}).\\

To find the explicit solution for the \emph{cone} case, the authors present the following map (see equation 4.8 of Cottenden \emph{et al.} \cite{cottenden2009analytical}),
\begin{align}
\boldsymbol\sigma(x,y) &= \frac{r}{\sqrt{1 + R(\phi(\theta))^2}}\boldsymbol(R(\phi(\theta))\cos(\phi(\theta)), ~R(\phi(\theta))\sin(\phi(\theta)), ~1\boldsymbol) ~,\label{dave6.1}
\end{align}
where
\begin{align*}
d\theta & = \frac{\sqrt{R^2 + (R^\prime)^2 + R^4}}{1 + R^2}d\phi = \frac{R\sqrt{1 + R^2}}{\sqrt{(1 + R^2)^2 - (R_{,\theta})^2}}d\phi~,\\
r &= \sqrt{x^2+y^2} ~~ \text{and}\\
\theta & = \arctan(\frac{y}{x}) - \zeta~.
\end{align*}
From the authors' definition, $\zeta$ appears to be the acute angle that the vector $\hat{\boldsymbol \sigma}_{\!,y}$ makes with the vector $\hat {\boldsymbol \sigma}_{\!,\phi}$, $R$ is given as a `cylindrical polar function' and $\phi$ appears to be the angle of the centre of rotation. One can clearly see that map (\ref{dave6.1}) is only valid for right-circular cones (i.e. cones with a circular cross section) as it must have the same metric as the Euclidean plane. To be precise
\begin{align*}
\boldsymbol F_{I}=
&\begin{pmatrix}
{\boldsymbol \sigma}_{\!,x}\cdot {\boldsymbol \sigma}_{\!,x} & {\boldsymbol \sigma}_{\!,x}\cdot {\boldsymbol \sigma}_{\!,y}\\
{\boldsymbol \sigma}_{\!,x}\cdot {\boldsymbol \sigma}_{\!,y} & {\boldsymbol \sigma}_{\!,y}\cdot {\boldsymbol \sigma}_{\!,y}
\end{pmatrix}\\
= & \begin{pmatrix}
1  & \frac{(1 - R)}{1+R^2} (\phi^\prime R^\prime)^2\\
\frac{(1 - R)}{1+R^2} (\phi^\prime R^\prime)^2 & \frac{(\phi^\prime R^\prime)^2(1 - 2R)}{(1+R^2)^2} + \frac{(\phi^\prime)^2[R^2 +  (R^\prime)^2]}{1+R^2}
\end{pmatrix}\\
= &\begin{pmatrix}
1 &0\\
0 & 1
\end{pmatrix} ~ ,
\end{align*}
which, in turn, implies that $R^\prime =0$ (as $\phi^\prime \neq 0 $). Thus, equation (\ref{dave6.1}) reduces to the following form,
\begin{align*}
\boldsymbol\sigma(x,y) &= r \boldsymbol(\sin(\alpha) \cos(\phi(\theta), ~\sin(\alpha) \sin(\phi(\theta)), ~\cos(\alpha)\boldsymbol) ~,\\
\phi(\theta) & = \operatorname{cosec}(\alpha) \theta~, \nonumber
\end{align*}
which represents a parametrisation of a right-circular cone, where $2\alpha$ is the (constant-) angle of aperture. Now, given that a solution exists in the interval $[\theta_1,\theta_2]$, the authors states that the solution can be expressed as follows (see equation 4.20 of Cottenden \emph{et al.} \cite{cottenden2009analytical}),
\begin{align}
T_{yy}(\phi_2) = T_0\exp\left(\frac{\mu_d}{R(\phi(\theta))}\sin\left(\theta + \zeta\right) \big|^{\theta_2}_{\theta_1} \right)~. \label{dave5}
\end{align}

Regardless of the authors' claims, solution (\ref{dave5}) is only valid for right-circular cones, i.e. valid for $R = \tan(\alpha)$ where $2\alpha$ is the (constant-) angle of aperture. To see why equations (\ref{dave6.1}) and (\ref{dave5}) is invalid for a general cone, one only needs to consider an example with noncircular cross section. If the reader wishes to, then consider a positively-oriented elliptical-cone (for the $\zeta= 0$ case) that is defined by the following map,
\begin{align}
\boldsymbol\sigma(\phi,z) = \boldsymbol(a z \cos(\phi), ~b z \sin(\phi), ~z\boldsymbol)~, \label{coneX}
\end{align}
where $z \in \mathbb R_{>0}$  and $a,b>0$, and let $\phi \in [\frac14\pi,\frac34\pi]$ be the contact interval (see figure \ref{treck1} and consider the $z=1$ case). Now, the reader can that  both map (\ref{dave6.1}) and solution (\ref{dave5}) is no longer valid for the elliptical-cone mapping (\ref{coneX}).\\

The authors conclude by stating that their experimental results agreed almost perfectly with equation (\ref{dave4}) for the cylinder case. One expects that the solution to agree with experiment data for the cylinder case as the solution is only valid for the cylinder case. The authors further state the `Experimental data gathered on [right-circular] cones constructed from plaster of Paris and Neoprene ... with half-angles ranging up to $12^\circ$ and contact angles in the range $[70^\circ, 120^\circ]$ show good agreement with the simple cylindrical model at their error level (around $\pm10\%$ for most samples)' \cite{cottenden2009analytical}. Again, one expects this be the case as solution (\ref{dave5}) is only valid for right-circular cones. Also, it is given by the authors that in the limit $R \rightarrow 0$, the cone case is asymptotically equals to the cylinder case. This result is just a trivial mathematical result that follows directly from the Maclaurin series, i.e. $\sin(\theta) \approx \theta $, when $\theta \approx 0$.\\

In Cottenden \emph{et al.}'s publication \cite{cottenden2009analytical}, the authors fail to demonstrate a sufficient knowledge in the subject of differential geometry, mathematical elasticity and contact mechanics to tackle this problem with any mathematical rigour, and this is evident in D. J. Cottenden's  thesis \cite{cottenden2011multiscale} as the publication Cottenden \emph{et al.} \cite{cottenden2009analytical} is a summary of all the mathematical results from Cottenden's  thesis \cite{cottenden2011multiscale}. For example, in section 2.15 of the thesis, the compatibility conditions for the left Cauchy-Green deformation tensor is given as a general condition (see page 8 of Cottenden \cite{cottenden2011multiscale}). However, there exists no general compatibility condition for the left Cauchy-Green deformation tensor, and the given compatibility conditions exist for the two-dimensional case only \cite{acharya1999compatibility}. Another example is that the entire section 5.4 of the thesis (see pages 132 to 137 of Cottenden \cite{cottenden2011multiscale}) is based on the assumption that one can invert a $3\times 2$ matrix (see equation 5.15 of Cottenden \cite{cottenden2011multiscale}), i.e. given a sufficiently differentiable 2D manifold $\boldsymbol \lambda :\mathbb R^2 \to \textbf{E}^3$ (e.g. a map of a cylinder), the author asserts that  the Jacobian matrix, $\boldsymbol ( \partial_\beta \lambda^j \boldsymbol )_{3\times 2}$ (where $\beta \in\{1,2\}$ and $j\in\{1,2,3\}$), is invertible: note that the author is considering regular inverse and not one-sided inverse. Should the reader consult section 5.4.1 of Cottenden \cite{cottenden2011multiscale}, it is evident that the author failed to understand the difference between the inverse of a bijective mapping and a preimage (which need not be bijective), as both definitions expressed with the same mathematical notation, $\boldsymbol \lambda^{-1}$, in Pressley's publication \cite{pressley2010elementary}: recall that Cottenden \cite{cottenden2011multiscale} accredits Pressley \cite{pressley2010elementary} for his differential geometry results. This misunderstanding of Pressley's work \cite{pressley2010elementary} leads to a substantial part of Cottenden's work \cite{cottenden2011multiscale} being incorrect, as Section 5.4 of Cottenden's thesis \cite{cottenden2011multiscale} is based on Cottenden's  assumption \cite{cottenden2011multiscale} that a $3\times 2$ matrix is invertible (where the author is considering regular inverse).\\

In a different publication, a precursor to the one we discussed, Cottenden \emph{et al.} \cite{cottenden2008development} present experimental framework to calculate the coefficient of friction, based on the master's thesis of Skevos-Evangelos Sp. Karavokyros\footnote{\href{https://liveuclac-my.sharepoint.com/:b:/g/personal/zcahe58_ucl_ac_uk/EQ_ktlQ5uhBGsZEOeY797SwBb02EPRbiyVUdVN6_T_tupg}{\emph{Validating a mathematical model with a simple laboratory model}. MSc. UCL. 2017.}}. The authors state that `The model generalizes the common assumption of a cylindrical arm to any convex prism' \cite{cottenden2008development}. Coefficients of friction are determined from experiments conducted on Neoprene-coated Plaster of Paris prisms of circular and elliptical cross-sections (defined as arm phantoms) and a nonwoven fabric. The authors state experimental results agreed within $\pm8\%$, i.e. $16\%$. They also state that the coefficients of friction varied very little with the applied weight, geometry and contact angle. Thus, the authors conclude by asserting that accurate values of the coefficient of friction can be obtained by applying the cylindrical model (i.e. the capstan equation) to the experimental data on human arms. They further assert that the coefficient of friction is independent of the substrate geometry, applied weights and contact angles, and claims that both their mathematical model and experimental results are in complete agreement.\\

Unfortunately, none of Cottenden \emph{et al.}'s  mathematical results \cite{cottenden2008development} can be mathematically justified, mostly for the reasons that we described before. For example, the reader may try to derive an arc-length of an ellipse with the use of the definition of an arc length from section 2.4 of the publication: although, the formula $\text{d}[\text{arc length}] = \sqrt{R(\theta)^2 + \frac{\text{d}R(\theta)}{d\theta} }\text{d}\theta$ holds true when calculating an arc-length of a curve (which can be derived with simple differential geometry techniques), the term $\text{d}[\text{arc length}]^2 = (R\text{d}\theta)^2 + \text{d}R^2$ (see directly above equation 12 of Cottenden \emph{et al.} \cite{cottenden2008development}) does not imply the former equation nor does it have any mathematical context. Another example is the equation $1/\tan(0.5\pi) = 0$, which is from the latter part of section 4.1 of the publication  (see directly below equation 17 of Cottenden \emph{et al.} \cite{cottenden2008development}).\\

\begin{figure}[!h]
\centering
\includegraphics[width=\linewidth]{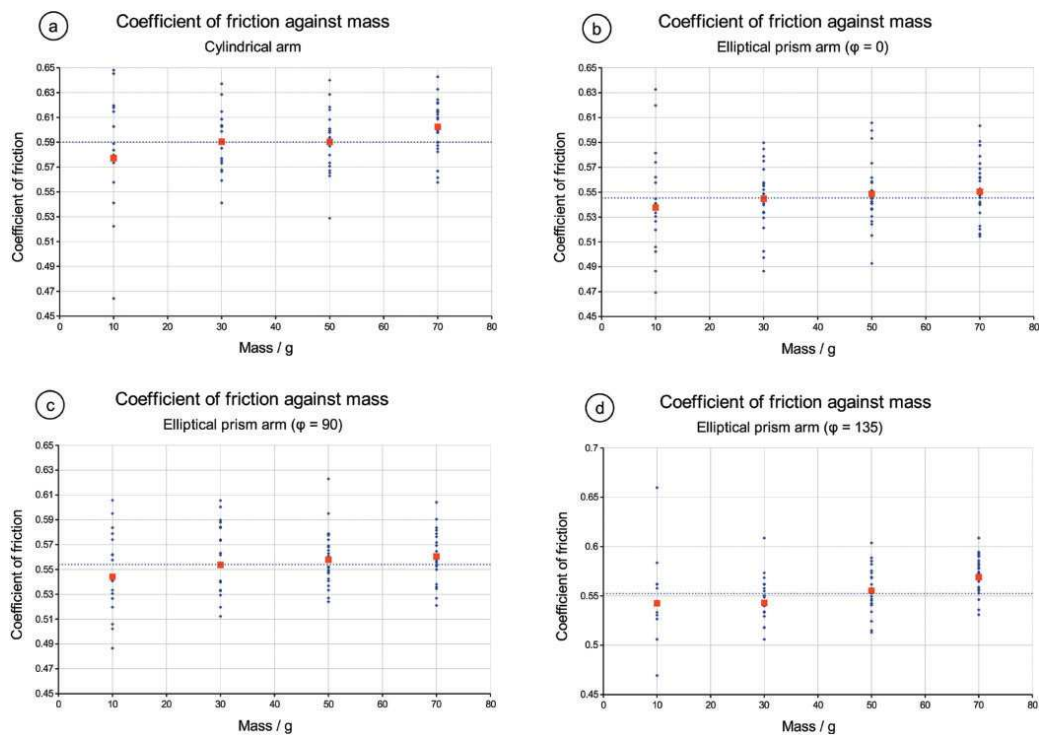}
\caption{Coefficient of friction against applied mass in grams: (a) Cylindrical body; (b) Elliptical prism with horizontal major axis; (c) Elliptical prism with vertical major axis; (d) Elliptical prism with major axis making $+135^{\circ}$ to the horizontal \cite{cottenden2008development}. \label{David2008} }
\end{figure}

As for the experimental results, consider figure \ref{David2008}, which shows the coefficients of friction in relation to different geometries, applied weights, and contact angles (see figure 11 of Cottenden \emph{et al.} \cite{cottenden2008development}). One can see that there are clear discrepancies between each calculated coefficients of friction as the coefficients of friction vary between different geometries, weights, and contact angles. The authors only acknowledge the dependence of coefficient of friction relative to the applied weight (see section 4 of Cottenden \emph{et al.} \cite{cottenden2009analytical}), but hastily dismiss this effect by asserting that `... the increase [coefficient of friction relative to the applied weight] is small compared to the scatter in the data' \cite{cottenden2008development}. \\

\begin{table}[!h]
\begin{center}
\begin{tabular}{ |c||r|r|r|r|r|c| }
\hline
\multirow{2}{*}{Mass g} &\multicolumn{6}{ |c| }{ Tension $10^{-3}$N} \\
\cline{2-7}
& 1st & 2nd & 3rd & 4th & 5th & Mean \\ \hline\hline
$10$ & $16.0g$ & $15.0g$ & $15.0g$ & $15.0g$ & $16.0g$ & $15.6g$ \\ \hline
$30$ & $51.0g$ & $54.0g$ & $51.0g$ & $50.0g$ & $51.0g$ & $51.4g$ \\ \hline
$50$ & $88.0g$ & $87.0g$ & $89.0g$ & $87.0g$ & $90.0g$ & $88.2g$ \\ \hline
$70$ & $125g$ & $124g$ & $128g$ & $122g$ & $124g$ &$125g$ \\ \hline
\end{tabular}
\caption{Tensometer readings: Cylinder with $\frac{127}{360}\pi$ contact angle, where $g$ is the acceleration due to gravity. \label{KaraCylinder}}
\end{center}
\end{table}

\begin{figure}[!h]
\centering
\includegraphics[width=0.75\linewidth]{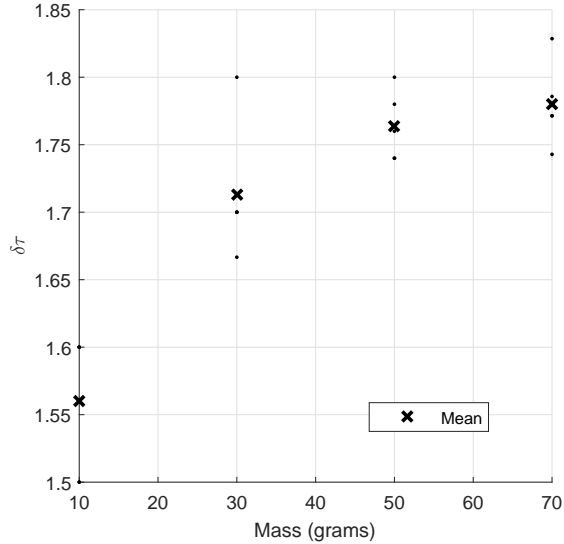}
\caption{Tension ratio against applied mass. \label{DeltaTau}}
\end{figure}

If one consults \href{https://liveuclac-my.sharepoint.com/:b:/g/personal/zcahe58_ucl_ac_uk/EQ_ktlQ5uhBGsZEOeY797SwBb02EPRbiyVUdVN6_T_tupg}{Karavokyros' masters thesis} for the experimental data, then one can find the raw data for the cylinder, $\frac{127}{360}\pi$ contact angle case (see table 2a of  \href{https://liveuclac-my.sharepoint.com/:b:/g/personal/zcahe58_ucl_ac_uk/EQ_ktlQ5uhBGsZEOeY797SwBb02EPRbiyVUdVN6_T_tupg}{Karavokyros' masters thesis}), which is displayed in table \ref{KaraCylinder}. Now, using this data, if one plot the tension ratio against the applied mass, then one gets figure \ref{DeltaTau}. Note that the capstan equation implies that the tension ratio is constant for all applied masses, i.e. $\delta \tau = \exp(\mu_d\theta_0)$ is constant given that $\mu_d$ and $\theta_0$ are constants. However, this is not what the experimental results are implying, as the reader can clearly see from figure \ref{DeltaTau} that as the applied mass increases, tension ratio also increases, and this is documented phenomenon in the literature \cite{jung2008capstan}, which cannot be simply dismissed. Thus, this implies that, for the given experiments, it is unsuitable to use the standard capstan equation to find the coefficient of friction with a significant degree of accuracy. Now, this is direct evidence that shows the flaws in Cottenden \emph{et al.}'s \cite{cottenden2009analytical} data analysis methods and their interpretation of the experimental results.\\

Unfortunately no raw data is available for the other experiments in the theses of \href{https://liveuclac-my.sharepoint.com/:b:/g/personal/zcahe58_ucl_ac_uk/EQ_ktlQ5uhBGsZEOeY797SwBb02EPRbiyVUdVN6_T_tupg}{Karavokyros} or Cottenden \cite{cottenden2011multiscale} to conduct further rigorous analysis, as we did with the $\frac{127}{360}\pi$-cylinder case.\\

As a result of the flawed mathematics and data analysis techniques, Cottenden \emph{et al.} \cite{cottenden2008development, cottenden2009analytical} and Cottenden \cite{cottenden2011multiscale} assert that the tension observed on the membrane is independent of the geometry and the elastic properties of the foundation, and thus, the stress profile at the contact region and the coefficient of friction can be calculated  with the use of the ordinary capstan equation. They further assert that the experimental methodology for calculating the coefficient of friction between fabrics and in-vivo (i.e. within the living) human skin is identical to the capstan model. However, we found no experimental evidence in the body of the authors' publications to verify their assertion, i.e. no evidence is given for the assertion that foundation's geometric and elastic properties are irrelevant when calculating the coefficient friction between an elastic foundation and an overlying membrane. Thus, our next subject of investigation is the authors' experimental methodology.

\section{Experimental Data: June 2015 Human Trial \label{BeltExperiments}}

We now analyse the data obtained from human trials based on Cottenden \emph{et al.} \cite{cottenden2008development, cottenden2009analytical} and Cottenden's \cite{cottenden2011multiscale} experimental methodology. We recruit $10$ subjects, $5$ males and $5$ females, all between the ages of $18$ to $60$, and the approval was granted by the UCL Research Ethics Committee: UCL Ethics Project ID number 5876/001. Collected data of the subjects can be  found in table \ref{Experimenteedata}, if the reader wishes to reproduce any results, where BMI is the body mass index (calculated with NHS BMI calculator \footnote{\href{https://www.nhs.uk/live-well/healthy-weight/bmi-calculator/}{https://www.nhs.uk/live-well/healthy-weight/bmi-calculator/}}), Radius is the radius of the upper arm and  $\delta l$ is a measure of how \emph{flaccid} the subject's tissue is (see equation (\ref{flacid})). For more comprehensive set of raw data, please consult Dr N. Ovenden (UCL) at n.ovenden@ucl.ac.uk.\\

\begin{table}[!h]
\begin{center}
\begin{tabular}{ |c||c|c|c|c|c|}
\hline
Subject & Gender & Age (Years) &BMI& Radius (cm) & $\delta l$\\ \hline\hline
F19& Female& $19$& $21.0$& $3.98$ &$0.994$\\ \hline
F34& Female& $34$& $22.0$& $4.22$ &$0.991$\\ \hline
F40& Female& $40$& $23.4$& $3.82$ &$1.01$\\ \hline
F53& Female& $53$& $27.3$& $4.54$ &$1.02$\\ \hline
F60& Female& $60$& $22.5$& $4.46$ &$1.02$\\ \hline
M18& Male& $18$& $17.5$& $3.50$ &$0.98$\\ \hline
M23& Male& $23$& $24.7$& $4.77$ &$1.04$\\ \hline
M25& Male& $25$& $22.6$& $4.22$ &$1.01$\\ \hline
M26& Male& $26$& $22.8$& $4.50$ &$0.988$\\ \hline
M54& Male& $54$& $26.2$& $5.09$ &$1.00$\\ \hline
\end{tabular}
\caption{Experimentee data 2015. \label{Experimenteedata}}
\end{center}
\end{table}

\begin{figure}[!h]
\centering
\includegraphics[trim = 0cm 0cm 0cm 0cm, clip = true, width=0.825\linewidth]{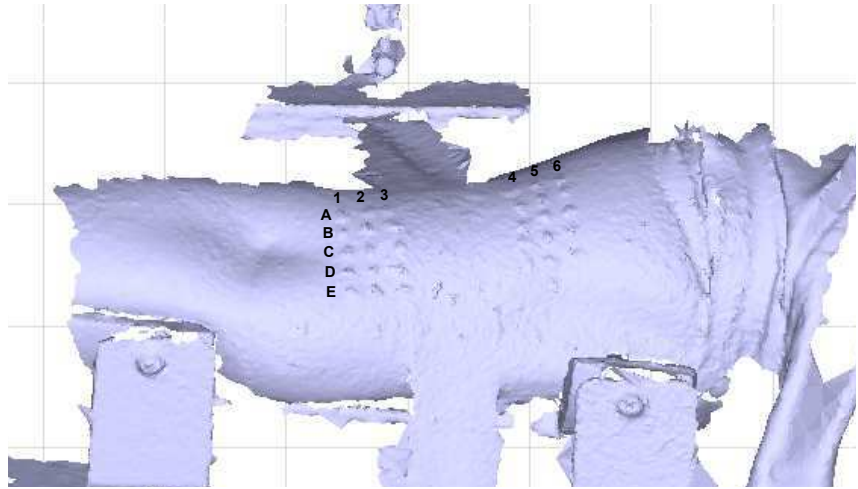}
\caption{Experimental configuration on subject F53 (.stl file). \label{Key}}
\end{figure}

For our experimental configuration, we place subjects upper arm horizontally, bicep facing upwards, on custom-designed  armrest. Then, we place a fabric (a nonwoven fabric: $95\%$ polypropylene and $5\%$ cotton) over their bicep and attach the upper end to the tensometer (Dia-Stron MTT170 provided by Dr S. Falloon, UCL), and the free hanging end is reserved for hanging set of weights, such that the contact angle between the bicep and the fabric (if measured from the humerus) is approximately $\frac12 \pi$. The dimensions of the fabric are $4$cm$\times 50$cm, and from our measurements, the fabric has an approximate thickness of $0.5$mm and an approximate mass of $0.6$g. Also, we mark the skin with a $3$cm$\times5$cm grid with $1$cm$\times1$cm grid spacing, $0.5$cm away from either side of the fabric and starting from the highest part (of the horizontal axis) of the arm, then placing semi-hemispherical markers with a radius of $2$mm. See figure \ref{Key} for a visual representation.\\

For each run, we pull the fabric with a constant speed of $\frac{1}{6}\text{cms}^{-1}$ with the use the tensometer and  we use a static-3D scanner (3dMD Photogrammetric System\footnote{\href{http://www.3dmd.com}{http://3dmd.com}} provided by Mr C. Ruff, UCLH) to record the before and after deformation of the upper arm. Note that we use the metric 
\begin{align}
\delta l = \frac{\sum_{10\text{g}, \dots,140\text{g}}||\text{deformed}(\textbf{A4}-\textbf{E6})||}{\sum_{10\text{g}, \dots,140\text{g}}||\text{undeformed}(\textbf{A4}-\textbf{E6})||}~,\label{flacid}
\end{align}
to measure the \emph{flaccidity} (analogous to $1/(\text{Young's modulus})$) of the subject's soft tissue. Also, Radius (from table \ref{Experimenteedata}) is calculated by measuring the girth around the bicep and then dividing the measurement by $2\pi$\\

\begin{table}[!h]
\begin{center}
\begin{tabular}{ |c||c|c|c|c|c|c|c|}
\hline
\multirow{2}{*}{Subject} &\multicolumn{6}{ |c| }{$\delta \tau$} \\
\cline{2-7}
& $40$g & $60$g & $80$g & $100$g & $120$g & $140$g \\ \hline\hline
F19& $2.19$& $1.70$& $1.52$& $1.92$& $1.88$& $1.86$\\ \hline
F34& $2.08$&$\cdots$&$1.93$& $1.89$& $1.83$& $1.83$\\ \hline
F40& $1.99$& $1.96$& $1.92$& $1.89$& $1.88$& $1.79$\\ \hline
F53& $2.13$& $2.31$& $2.23$& $2.15$& $1.84$& $1.70$\\ \hline
F60& $2.29$& $2.20$& $2.06$& $1.99$& $1.96$& $1.95$\\ \hline
M18& $1.99$&$1.90$& $1.88$& $1.84$& $1.68$& $1.76$\\ \hline
M23& $2.46$&$2.28$& $2.24$& $2.19$& $2.33$& $1.99$\\ \hline
M25& $2.14$&$1.98$& $1.81$& $1.80$& $1.98$&$\cdots$\\ \hline
M26& $2.41$&$2.31$& $2.26$& $2.18$& $2.31$& $1.99$\\ \hline
M54&$\cdots$&$2.12$&$2.03$& $1.77$& $1.91$& $1.96$\\ \hline
\end{tabular}
\caption{Experimental data from the 2015 human trial. \label{ParticipantFriction}}
\end{center}
\end{table}

Table \ref{ParticipantFriction} shows the tension ratio, $\delta\tau = T_\text{max}/T_0$, for each subject with respect to each applied mass, where $T_\text{max}$ are the tensometer readings, $T_0 = \text{Mass}\times g$ are the weight of the applied mass and $g =9.81$ is the acceleration due to gravity. Note that the tensometer data of F34 for $60$g, M25 for $140$g and M54 for $40$g are corrupted, and thus, omitted.\\

Cottenden \emph{et al.} \cite{cottenden2008development, cottenden2009analytical} and Cottenden \cite{cottenden2011multiscale} assert  that the $T_\text{max} = T_0\exp(\mu_F\theta_0)$ equation holds true, regardless of the geometric and elastic properties of the substrate (i.e. human soft-tissue). If this is indeed true then $\delta\tau= \exp(\mu_F\theta_0)$ is only a function of the coefficient of friction,  regardless of the geometric and elastic properties of human soft-tissue, for a fixed contact angle $\theta_0$. Also, one of the major assumption of the authors is that coefficient of friction between skin and fabrics positively correlated with the age of an individual, as they observe higher occurrence of skin damage in elderly subjects. Thus, now we plot the tension ratio against various properties to test the authors' claims.\\

\begin{figure}[!h]
\centering
\begin{subfigure}{.5\linewidth}
    \centering
    \includegraphics[width=1\linewidth]{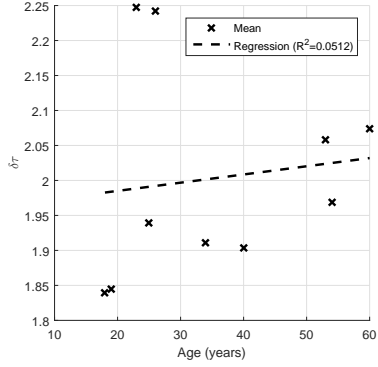}
\caption{Tension ratio against the age (in years).\label{PAge}}
\end{subfigure}%
\begin{subfigure}{.5\linewidth}
    \centering
    \includegraphics[ width=1\linewidth]{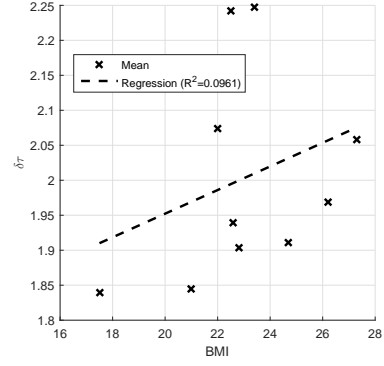}
\caption{Tension ratio against the BMI. \label{PBMI}}
\end{subfigure}
\begin{subfigure}{.5\textwidth}
    \centering
    \includegraphics[width=1\linewidth]{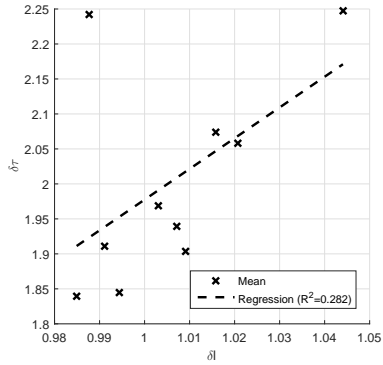}
\caption{Tension ratio against the $\delta l$. \label{PL}}
\end{subfigure}%
\begin{subfigure}{.5\textwidth}
    \centering
    \includegraphics[width=1\linewidth]{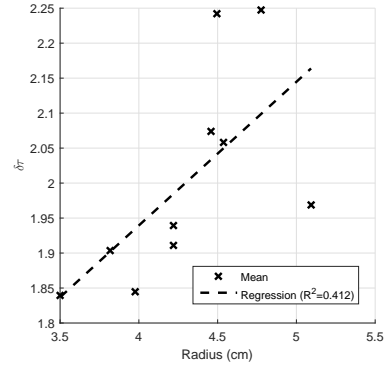}
\caption{Tension ratio the radius of the upper arm (in cm). \label{PR}}
\end{subfigure}
\caption{Tension ratio for varying age,  BMI, $\delta l$ and radius of the upper arm.}
\end{figure}

Figure \ref{PAge} shows the tension ratio against subjects' age, where the linear regression line is $\delta\tau= 0.00117 \times\text{Age} + 1.96$ and the age is in years ($\text{R}^2 = 0.0512$, where $R^2$ is the coefficient of determination). As the reader can see there is no obvious relationship between the age of the subject and the tension ratio. In fact, the highest tension ratios are observed for M23 and M26 (two of the youngest subjects).\\

Figure \ref{PBMI} shows the tension ratio against the body mass index (BMI), where the linear regression line is $\delta\tau= 0.0169\times\text{BMI} + 1.61$ ($\text{R}^2 = 0.0961$). As the reader can see that there is a vague positive correlation between the BMI and the tension ratio. The reason we observe this correlation is because that those who have a higher BMI tend to have a greater fat content, i.e. higher volume of flaccid tissue. Thus, during the experiments, a higher tension needs to be applied to the fabric as the as a portion of the applied-tension is expended on deforming the flaccid tissue of the subject.\\

Figure \ref{PL} shows the tension ratio against $\delta l$, where the linear regression line is $\delta\tau= 4.39\times\delta l- 2.41$ ($\text{R}^2 = 0.282$). As the reader can see that there is a positive correlation between $\delta l$ and the tension ratio. This implies that for more flaccid  soft-tissue, larger portion of the applied-tension is expended on deforming it. \\

Figure \ref{PR} shows the tension ratio against the radius of the upper arm, where the linear regression line is $\delta\tau= 0.205\times\text{Radius} + 1.12$ and the radius is in centimetres ($\text{R}^2 = 0.412$). As the reader can see that see there is a strong positive correlation between the radius and the tension ratio. The reason why we observe this correlation is exactly same as the cases before: the larger the radius is, the larger the volume of soft-tissue that needs to be deformed, and thus, a larger applied-tension.\\

From figures \ref{PBMI}, \ref{PR} and \ref{PL}, we observe higher tension ratios for subjects with higher BMI, more flaccid soft-tissue, and larger biceps. Should one use the capstan equation (or any belt-friction model in general) to calculate the coefficient of friction, then one would observe a higher coefficient of friction for subjects with higher BMI, more flaccid soft-tissue, and larger biceps. However, this does neither imply nor does not imply that those with a higher BMI, more flaccid soft-tissue, and larger biceps have a greater risk of skin abrasion, i.e \emph{correlation does not imply causation} \cite{aldrich1995correlations}; It merely implies that belt-friction models are not suitable for modelling such problems, which directly contradicts Cottenden \emph{et al.} \cite{cottenden2008development, cottenden2009analytical} and Cottenden's \cite{cottenden2011multiscale} assertion regarding the efficacy of the capstan model.\\

For a mathematically rigorous way of modelling this problem, we refer the reader to section 6.6 and 6.7 of Jayawardana \cite{jayawardana2016mathematical}. There, both numerical-modelling (shell-membrane frictionally coupled to an elastic-foundation) and human trial data imply that given a constant coefficient of friction, a higher volume of soft-tissue (high radius) and more compliant soft-tissue (lower Young's modulus) would result in higher deformation of the skin, and a higher volume of soft-tissue would result in more shear-stress generated on the skin.

\section{Conclusions}

In conclusion, we showed that no mathematical claim of Cottenden \emph{et al.} \cite{cottenden2008development, cottenden2009analytical} and Cottenden \cite{cottenden2011multiscale} can be mathematically justified, and some fundamental and trivial results in mathematical elasticity, differential geometry and contact mechanics are misrepresented. Only the ordinary capstan equation, and a solution to a dynamic membrane with both a zero-Poisson's ratio and a zero-mass density on a right-circular cone is given. Finally, limited experimental data is given to show the trivial asymptotic nature of $\sin(\theta)$ near $\theta=0$.\\

Also, Cottenden \emph{et al.} \cite{cottenden2008development, cottenden2009analytical} and Cottenden \cite{cottenden2011multiscale} claim that the capstan equation (\ref{CapstanEqn}) holds true, regardless of the geometric and elastic properties of the obstacle, and thus, it can be used to calculate the coefficient of friction between human skin and fabrics. However, the data gathered from human trials implies that it is unwise to use the capstan equation  (e.g. belt-friction models in general) to calculate the friction between in-vivo skin and fabrics. This is because that such models assume a rigid obstacle while human soft-tissue is elastic, and thus, a portion of the applied force to the fabric is expended on deforming the soft-tissue, which in turn leads to the illusion of a higher coefficient of friction when such models are used to calculate the coefficient of friction.

\bibliographystyle{./model1-num-names}
\bibliography{CriticaStudyOfCottendensWork}%CriticaStudyOfCottendensWork
\biboptions{sort&compress}

\end{document}